\documentclass[journal]{IEEEtran}
\IEEEoverridecommandlockouts
\usepackage{cite}
\usepackage{amsmath,amssymb,amsfonts}
\usepackage{amsthm,bm}
\usepackage{graphicx}
\usepackage{subfigure} 
\usepackage{textcomp}
\usepackage{url}
\usepackage{mathtools}
\def\BibTeX{{\rm B\kern-.05em{\sc i\kern-.025em b}\kern-.08em
    T\kern-.1667em\lower.7ex\hbox{E}\kern-.125emX}}

\usepackage{stackengine}
\usepackage{algorithm}
\usepackage{enumerate}
\usepackage{booktabs}
\usepackage{subfig}
\usepackage{multicol}
\usepackage[noend]{algpseudocode}
\usepackage[table,xcdraw]{xcolor}
\usepackage{pgfplots}
\usepackage{pgfplotstable}
\makeatletter
\def\BState{\State\hskip-\ALG@thistlm}
\makeatother

\algnewcommand\algorithmicforeach{\textbf{for each}}
\algdef{S}[FOR]{ForEach}[1]{\algorithmicforeach\ #1\ \algorithmicdo}
    
\begin{document}

\title{\huge Semantic Communication Meets Edge Intelligence}

\author{Wanting Yang, Zi Qin Liew, Wei Yang Bryan Lim, Zehui Xiong, \\Dusit Niyato, Xuefen Chi, Xianbin Cao, Khaled B. Letaief \thanks{W.~Yang is with the Department of Communications Engineering, Jilin University, Changchun, China, and also with Information Systems Technology and Design Pillar, Singapore University of Technology and Design, Singapore. Email: yangwt18@mails.jlu.edu.cn. 
ZQ.~Liew is with with Alibaba Group and the Alibaba-NTU Joint Research Institute, Nanyang Technological University, Singapore. Email: ziqin001@e.ntu.edu.sg.
WYB.~Lim is with Alibaba Group and Alibaba-NTU Joint Research Institute (JRI), Nanyang Technological University (NTU), Singapore. Email: limw0201@e.ntu.edu.sg. Z.~Xiong is with Information Systems Technology and Design Pillar, Singapore University of Technology and Design Singapore. Email: zehui\_xiong@sutd.edu.sg. D.~Niyato is with School of Computer Science and Engineering, NTU, Singapore. Emails: dniyato@ntu.edu.sg and ascymiao@ntu.edu.sg.
X.~Chi is with the Department of Communications Engineering, Jilin University, Changchun 130012, China. Email:  chixf@jlu.edu.cn.
X.~Cao is with School of Electronic and Information Engineering, Beihang University, Beijing, China. Email: xbcao@buaa.edu.cn.
K. B.~Letaief is with Department of Electronic and Computer Engineering, The Hong Kong University of Science and Technology (HKUST), Hong Kong. Email: eekhaled@ust.hk.}}

\makeatletter
\setlength{\@fptop}{0pt}
\makeatother

\maketitle

\begin{abstract}
The development of emerging applications, such as autonomous transportation systems, are expected to result in an explosive growth in mobile data traffic. As the available spectrum resource becomes more and more scarce, there is a growing need for a paradigm shift from  Shannon's Classical Information Theory (CIT) to  semantic communication (SemCom). Specifically, the former adopts a ``transmit-before-understanding" approach while the latter leverages artificial intelligence (AI) techniques to ``understand-before-transmit", thereby alleviating bandwidth pressure by reducing the amount of data to be exchanged without negating the semantic effectiveness of the transmitted symbols. However, the semantic extraction (SE) procedure incurs costly computation and storage overheads. In this article, we introduce an edge-driven training, maintenance, and execution of SE. We further investigate how edge intelligence can be enhanced with SemCom through improving the generalization capabilities of intelligent agents at lower computation overheads and reducing the communication overhead of information exchange. Finally, we present a case study involving semantic-aware resource optimization for the wireless powered Internet of Things (IoT).
\end{abstract}


\newtheorem{definition}{Definition}
\newtheorem{lemma}{Lemma}
\newtheorem{theorem}{Theorem}

\newtheorem{property}{Property}

\section{Introduction}

 With the ongoing convergence of information and communication technologies (ICTs) and artificial intelligence (AI), the ``Internet of Everything" has been considered as one of the key 6G visions, wherein \textit{semantic communication}  (SemCom) and \textit{edge intelligence} are expected to be two key enablers~\cite{letaief2021edge}.

SemCom is widely regarded as  a promising communication paradigm to breakout the ``Shannon's trap".  In fact, SemCom is not an entirely new concept.  Just after the introduction of Shannon's theorem,  Weaver and Shannon went on to identify three levels of problems within the broad subject of communication\cite{weaver1953recent}:
\begin{enumerate}
\item \textit{Technical level:} How accurately can the symbols of communication be transmitted? 
\item \textit{Semantic level:} How precisely do the transmitted symbols convey the desired meaning? 
\item \textit{Effectiveness level:} How effectively does the received meaning affect conduct in the desired way? 
\end{enumerate}
Shannon’s Classical Information Theory (CIT) focuses only on the technical level and achieves success in deriving a rigorous mathematical theory of communication based on probabilistic models, wherein the concept of information is defined as what can be used to remove uncertainty and the analysis is based on mutual information in the entropy domain. 

However, the achieved transmission rates in the CIT-driven conventional communication systems are approaching the Shannon limit and the available spectrum resources are becoming increasingly scarce. Moreover, the rapid development of emerging applications, e.g., autonomous transportation systems, leads to a never-ending growth in mobile data traffic. In this regard, SemCom has returned to relevance. Empowered by AI technologies such as computer vision (CV) and natural language processing (NLP), end devices such as sensor nodes or smartphones may eventually be equipped with human-like reasoning capabilities. In this way, \textit{semantic extraction} (SE) can be integrated into the communication model to achieve SemCom, which allows only the information of interest to the receiver 
for transmission, rather than raw data, thereby alleviating bandwidth pressure and enhancing privacy preservation by reducing the redundant data to be exchanged. However, there are still some factors hindering the implementation of SemCom. For instance, the training process of SE models requires significant computing and storage resources, thereby impeding the scalable implementation of SemCom on resource constrained end devices. Furthermore, in building a common knowledge base towards improving the generalization capabilities of SE models, other issues such as privacy loss may arise.



Fortunately, edge intelligence is promising to facilitate the scalable implementation of SemCom systems. The precursor to edge intelligence is edge computing, which moves part of the service-specific processing and data storage from the central cloud to the edge of the network closer to the source of data. In 5G networks, edge computing has already made significant achievements in terms of improving performance, and supporting new services and functions. Empowered by AI technologies  in 6G, edge intelligence 
aims to offer more powerful computational processing and massive data acquisition at the edge networks to achieve  the dynamic and adaptive edge maintenance and management~\cite{wang2020convergence}. Therefore, edge intelligence can provide a good basis for offloading SE model training and knowledge storage.

On the other hand, to realize the 6G vision of ubiquitous AI, \textit{distributed learning and inference} has become instrumental to contribute towards the intelligentization of edge networks~\cite{zhang2019mobile}. However, the data driven approach implies that AI enabled agents have to incur costly communication and computation overheads, which will pose challenges for communication network, especially amid the uncertain wireless environment and limited wireless resources. In this regard, SemCom can be seen as a key enabler of edge intelligence in turn.

Our contributions in this article is as follows:
\begin{itemize}
    \item We introduce a general system model for SemCom involving the three communication levels foreseen by Shannon and Weaver. We will then discuss some  performance metrics that differ from the CIT and some key SE techniques.
    
    \item To address the costly implementation overheads of training, maintaining, and executing SE models at the edge for SemCom, we introduce edge enabled SemCom by studying a Federated Learning (FL) enabled SE system model and edge-sharing knowledge graph for the semantic management. We will also introduce how SemCom can play a part in training semantic-aware intelligent agents and empowering communication efficient semantic-aware distributed machine learning.
    
    \item We insightfully discuss the open research issues and future research directions that are at the intersection of AI and communications toward SemCom, a key component of 6G networks.
    
    \item We provide a case study of semantic-aware resource allocation in wireless powered IoT. Different from the CIT driven IoT, our study utilizes an AI driven allocation mechanism to derive the resource allocation policy that maximizes semantic performance across the network.
\end{itemize}


\section{Preliminaries}

  SemCom differs from traditional Shannon communication in that it incorporates human-like ``understanding" and ``inference" into the encoding and decoding of communication data, no longer pursuing exact data replication. In this section, we provide a brief introduction of the SemCom framework and typical semantic metrics. We then discuss the key SE techniques in the existing works.

\subsection{SemCom Framework}
  \begin{figure}[t]
  \centering
  \includegraphics[scale = 0.35]{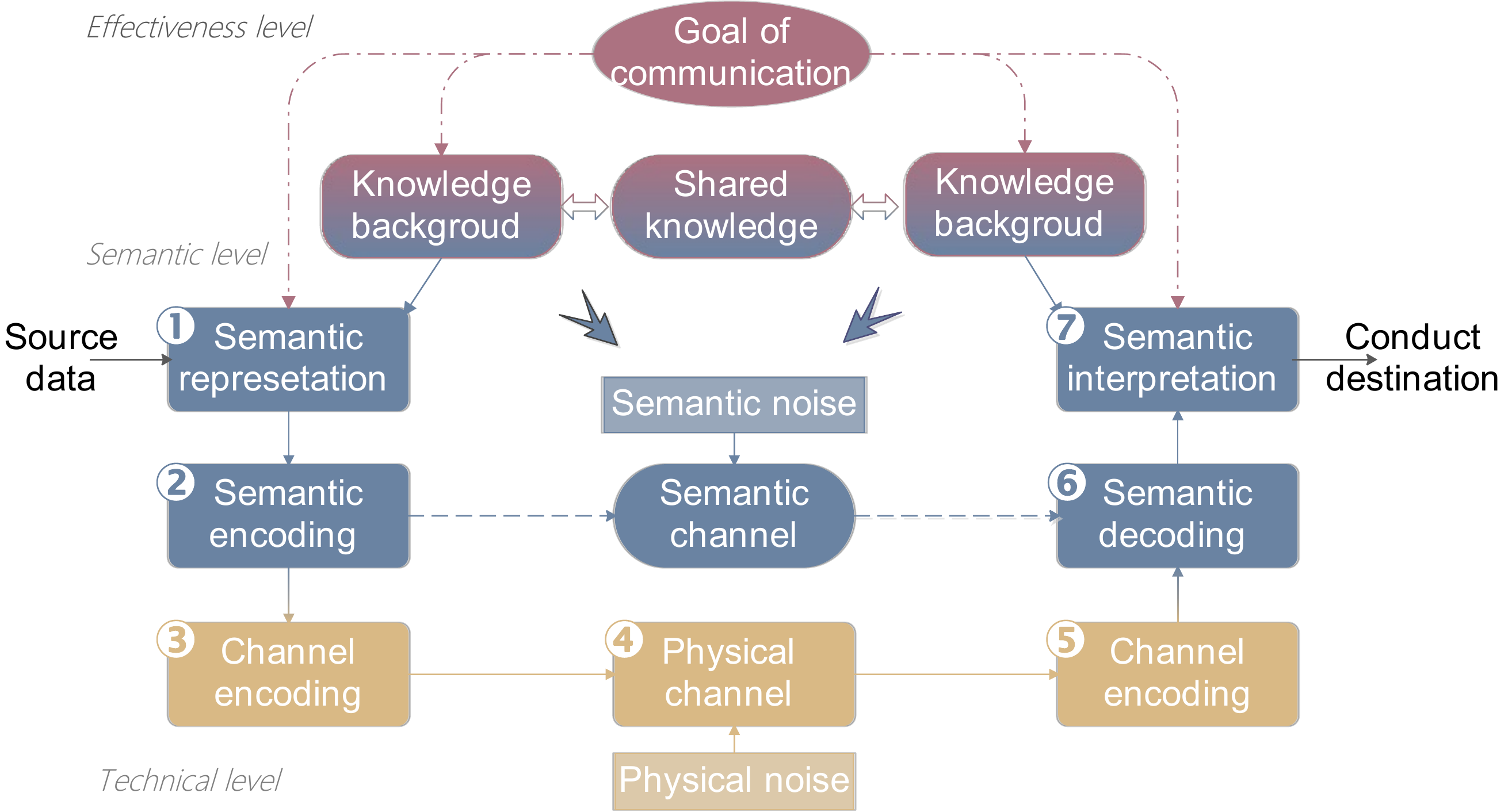}
  \caption{ SemCom model~\cite{yang2021semantic,weaver1953recent}.}
  \label{model}
\end{figure}

 Different from the content-blind classical communication systems, what matters in SemCom design is the accuracy of semantic content of source data, instead of the average information associated to the possibilities of source data that can be emitted by a source~\cite{strinati20216g}. As such,  the main changes in the SemCom system lie in the data processing phase before sending and after receiving.  (Fig.~\ref{model}). Before encoding, the source data  goes through the  semantic representation module, which can be seen as the ``understanding-before-transmission" process during which the  redundant information is removed. Then, the extracted useful and relevant information will go to the  semantic encoding module.
 In general SemCom scenarios,  semantic decoding is the inverse process of encoding, which are jointly determined based on the AI technologies and their  prior knowledge. For brevity, we refer to both semantic encoding and decoding as SE, and semantic encoding (decoding) in the subsequent text is considered to be integrated with the semantic representation (interpretation) module. 
 
As with human conversation, effective conversation requires common knowledge of each other's language and communication context. In SemCom, the \textit{background knowledge} of the communication parties has to be shared in real time to ensure that the processes of understanding and inference can be well-matched for all the source data. If the background knowledge fails to match, \textit{semantic noise} is generated, thereby resulting in performance degradation even in the absence of syntactic errors during the  physical transmission.
 Moreover, in some cases wherein the goal of communication may  change, all possibilities for communication goal should be included into the background knowledge and the communication goal should instruct SE to filter out irrelevant semantic information (SI) in each transmission.
 
 
\begin{table*}
\centering
\caption{Some semantic metrics  derived from NLP~\cite{lu2021reinforcement}}
\label{metrics}
\begin{tabular}{p{3cm} p{6cm} p{3cm} p{3cm}} 
\toprule
\multicolumn{2}{l}{\textbf{Semantic metrics}}& Advantages & Drawbacks                                                                                                                                                            \\ 
\hline
Bilingual evaluation understudy (BLEU)            & BLEU is a method for automatic evaluation for machine translation. It is used to is to compare word groups with different size of the candidate with that of the reference translation and count the number of matches. & It consider the linguistic laws, such as that semantically consistent words usually   come together in a given corpus.  &
It only calculates the differences of words between two sentences and has no insight into the meaning of the whole sentence.\\
\hline
Consensus based Image Description Evaluation (CIDEr) &  CIDEr was proposed as an automatic consensus metric of image description quality in, which was originally used to measure the similarity of a generated sentence against a set of ground truth sentences written by humans. It can also be used to evaluate the communication performance for text transmission. & Compared to BLEU, it does not evaluate semantic similarity on the basis of a reference sentence, but a set of sentences with the same meaning.  & Similar to BLUE, it is also based on the comparisons between word groups, and the semantic similarity can only be captured at the word level.    \\
\hline
Sentence similarity                                  &      Sentence similarity  is calculated as the cosine similarity of the semantic features extracted bidirectional encoder representations from transformers (BERT). BERT is a fine-tuned word representation model, which employs a huge pre-trained model including billions of parameters used for extracting the SI.    &  The SI in this metric is viewed from a sentence level owing to the sensitivity of BERT to polysemy, (e.g., word``mouse" in biology and machine).  & The pre-trained BERT model introduces much resource consumption in the training process and makes it hard to generalize in other tasks.                                                                                                                                                                                                          \\
\bottomrule
\end{tabular}
\end{table*}

  \begin{figure*}[t]
  \centering
  \includegraphics[width=\linewidth]{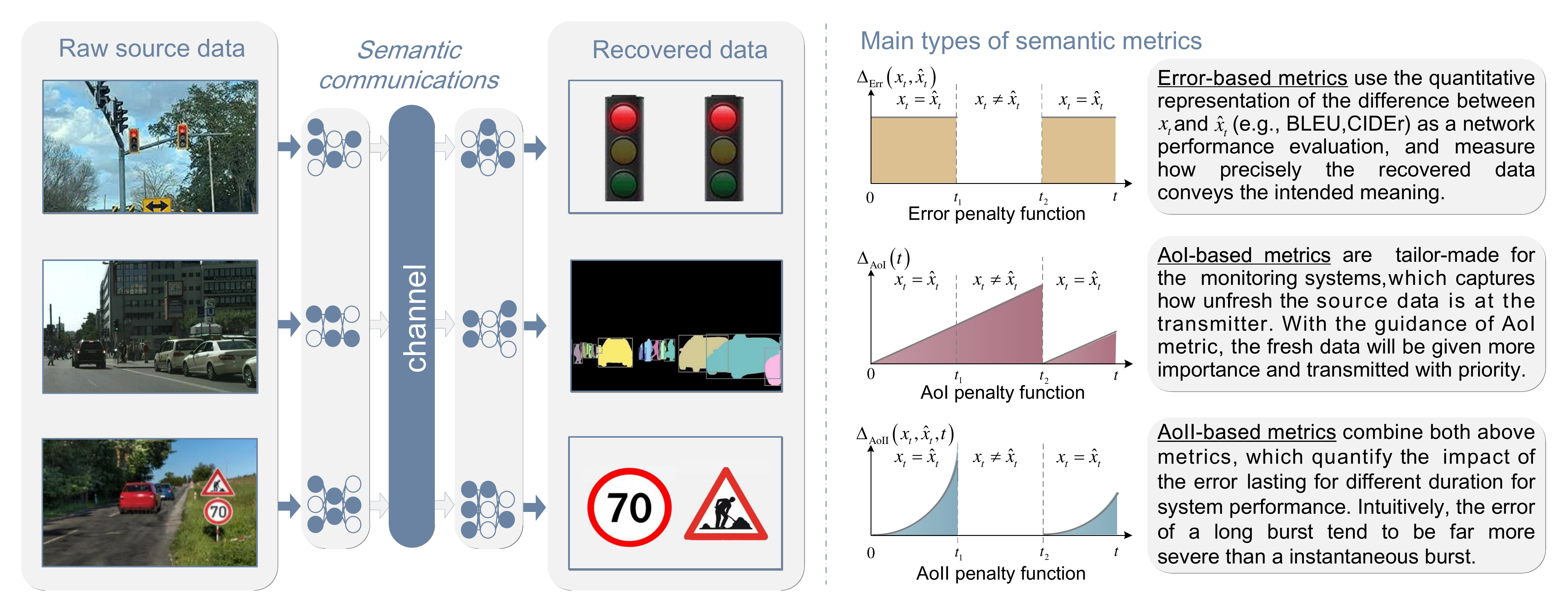}
  \caption{Some SemCom examples and metrics, where $x_t$ denotes the transmitted information, and $\hat{x}_t$ denotes the estimated information inferred from the transmission~\cite{maatouk2020age}.}
  \label{aoigraph}
\end{figure*}

\subsection{Semantic metrics}
The design of network performance metrics has long been a nucleus concern in network design and optimization. As the study on SemCom is still in its infancy, most semantic metrics are derived from NLP (Table~\ref{metrics}). Different from  bit-error rate (BER) or symbol-error rate (SER) in classical communication systems, they avoid treating packets equally, and measure the differences in the meaning conveyed by the recovered sentence and transmitted sentence. Besides such error-based metrics, there are some metrics focusing on timeliness. In fact, taking age of information (AoI) into account in performance evaluation can be initial attempts at SemCom. AoI-based metrics highlight the importance of freshness of the data packets, which allows scheduling scheme based on AoI minimization to filter out the irrelevant  packets  given the bandwidth constraints. By jointly considering the accuracy and timeliness of information, the authors in~\cite{maatouk2020age} introduce the metric of age of incorrect information (AoII) into SemCom, which can measure the network performance by looking at the bigger picture of the packet's role in achieving the overall communication goal.
Moreover, for the cases where the benefits of the packet content to be transmitted are evaluated to be important for the system objective, the value of the information (VoI) is of more concern than accuracy. Hence, the VoI-related metrics for SemCom are drawing increasing attention in goal-oriented communications that could capture the importance, relevance, and priorities of packets. Some SemCom examples and the three typical types of metrics are presented in Fig.~\ref{aoigraph}.

\subsection{Semantic extraction techniques}
\begin{figure*}[t]
  \centering
  \includegraphics[scale = 0.46]{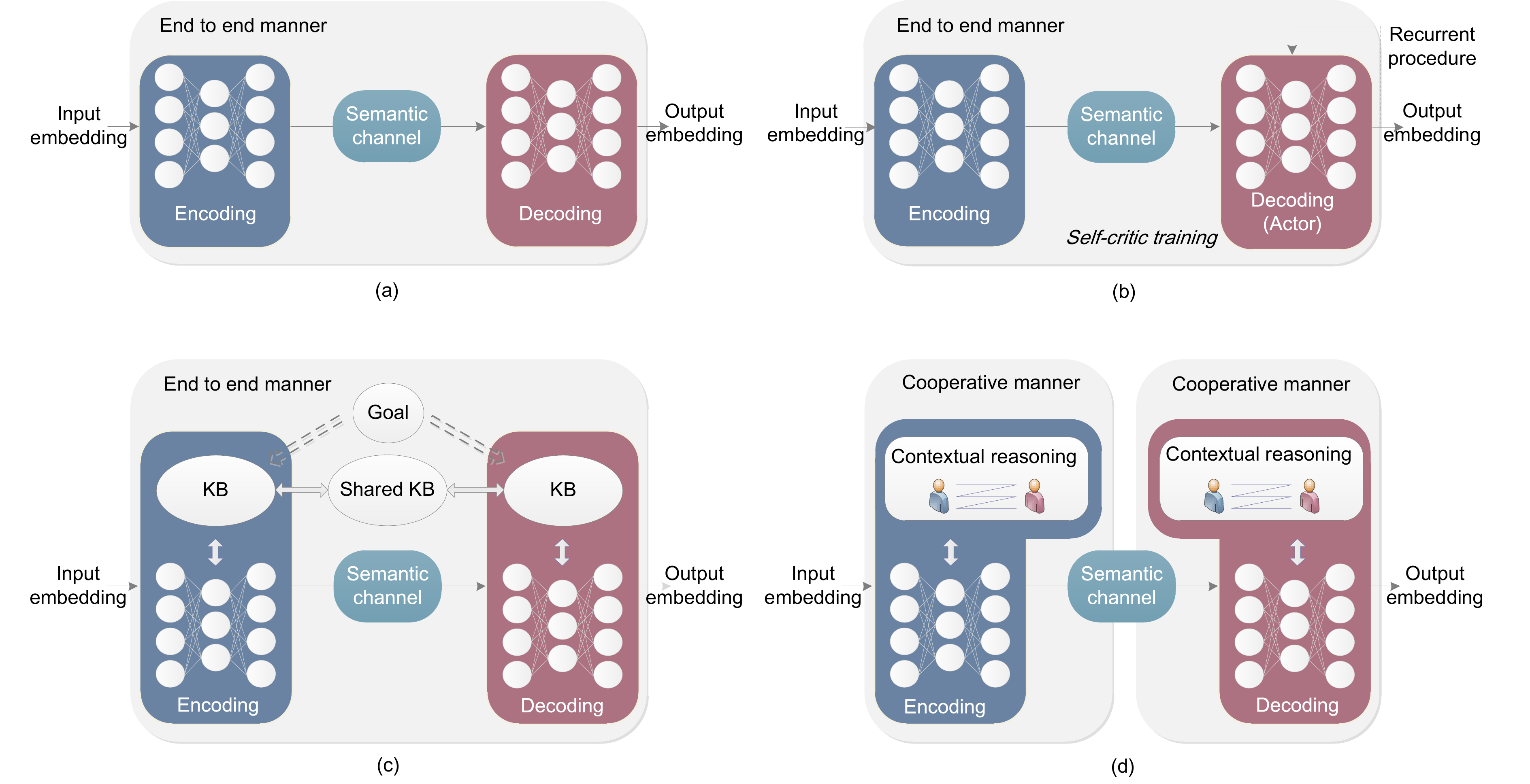}\\
  \caption{General models of main semantic extraction methods~\cite{xie2020lite,lu2021reinforcement,yang2021semantic,seo2021semantics}: a) system model of DL-based SE; b) system model of RL-based SE; c) system model of KB-assisted SE; d) system model of semantic-native SE. 
  }
  \label{methods}
\end{figure*}

We now discuss some key SE techniques, the general models of which are shown in Fig.~\ref{methods}.
\subsubsection{Deep learning based SE}
With the advancement of Transformer, squeeze-and-excitation network, and deep residual network, deep learning (DL) has been widely employed in SE for text, speech and image transmission. DL based SE aims to enhance the system robustness at low signal-to-noise ratio (SNR) with shorter bit-flow. The encoder and decoder are usually modeled as two separate learnable sections at the transmitter and receiver, and  linked through a random channel, which are trained jointly~\cite{xie2020lite}. During the training process, the Generative Adversarial Networks (GANs) are commonly used to model the channel dynamics and noise. However, as the loss function is generally required to be differentiable, the common loss functions such as cross entropy are adopted during the model training process. This treats the semantic contribution of all bits with equal importance, which is inconsistent with human perception in fact.

\subsubsection{Deep Reinforcement Learning based SE}
Deep Reinforcement learning (DRL)  can integrate the non-differentiable semantic metrics like BLEU into SE training.  In the DRL-based SE for text transmission in~\cite{lu2021reinforcement}, long short-term memory networks are employed in the encoder and decoder. The state is defined as the recurrent state of the decoder and the previously generated words. The transition between the two adjacent states is determined by the next generated word, and the action of the DRL agent is to generate a new word, with the action space defined by the dictionary dimension. As the semantic metrics can only be used as the long-term return in DRL, self-critic training is employed to address the challenging issue for identifying the  intermediate rewards, i.e., the impact of each step on the semantics of the whole sentence. Moreover, for other non-sequential task,  the decoding process needs to be transformed into a recurrent procedure beforehand.

\subsubsection{Knowledge Base assisted SE}
Knowledge Base (KB) is an emerging technology  widely  used in automated AI systems to store the data with formal representation  allowing for inference. The KB-assisted SE  integrates the KB into the encoder and decoder, aiming to extract SI more efficiently for scenarios with multiple communication tasks~\cite{yang2021semantic}.  Specifically,  the KB in Semcom is composed of source information, goals of the tasks, and the possible ways of reasoning that can be understood, recognized, and learned by all the communication participants. During the SE process, the KB is employed to quantify the level of relevance of SI for different communication tasks and instruct SE to capture the SI that is closely related to the task in each transmission. Meanwhile, as KB-assist SE is in an end-to-end manner, the KBs at both sides need to be kept in synchronization. 

\subsubsection{Semantic-native SE}
In the aforementioned methods, the SI is fixed. In~\cite{seo2021semantics}, the authors propose a semantic-native SE, wherein the SI can be learnt from iterative communications between intelligent agents, which make it feasible to the cases where the semantics vary over time and in different contexts. Moreover, the communication parties can be empowered with the capability of \textit{contextual reasoning} about the semantics in the local
context of social interactions, which  makes  communication more efficient. Hence, it can  promote intelligentization of communication systems with  high degree of flexibility and efficiency.


In summary, the DL-based SE is the most used. The RL-based approach, while achieving better performance than  DL-based ones, comes at the cost of a huge computational resource consumption. Moreover, KB-based SE is only validated for image classification and semantic-native SE remains a theoretical proposition currently.
\section{Edge-enabled SemCom}
\label{edge-enabled-semcom}
In contrast to the classical \textit{transmission-before-understanding} communications, the \textit{understanding-before-transmission} paradigm of SemCom requires a shared knowledge background and computationally costly operations for SE model training and inference. This undoubtedly poses new challenges summarized as follows:
\begin{itemize}
\item[1)]  Limited computing power and energy constraint of the end devices results in long latency in training and updating of the SE model, thereby degrading communication reliability.
\item[2)]  Comprehensive knowledge sharing among end devices to improve an SE model is at the cost of bandwidth and privacy. On the other hand, incomplete knowledge sets reduce the generalization capabilities of AI-based SE.
\item[3)] Most SE methods are task-specific and trained separately, which is far from brain-like cognition and is computationally-inefficient due to the duplication of work.
\end{itemize}
To address the above challenges, we propose an \textit{edge-enabled  SemCom architecture} in this section.

 \begin{figure*}[t]
  \centering
  \subfigure[]{
  \centering
  \includegraphics[scale = 0.35]{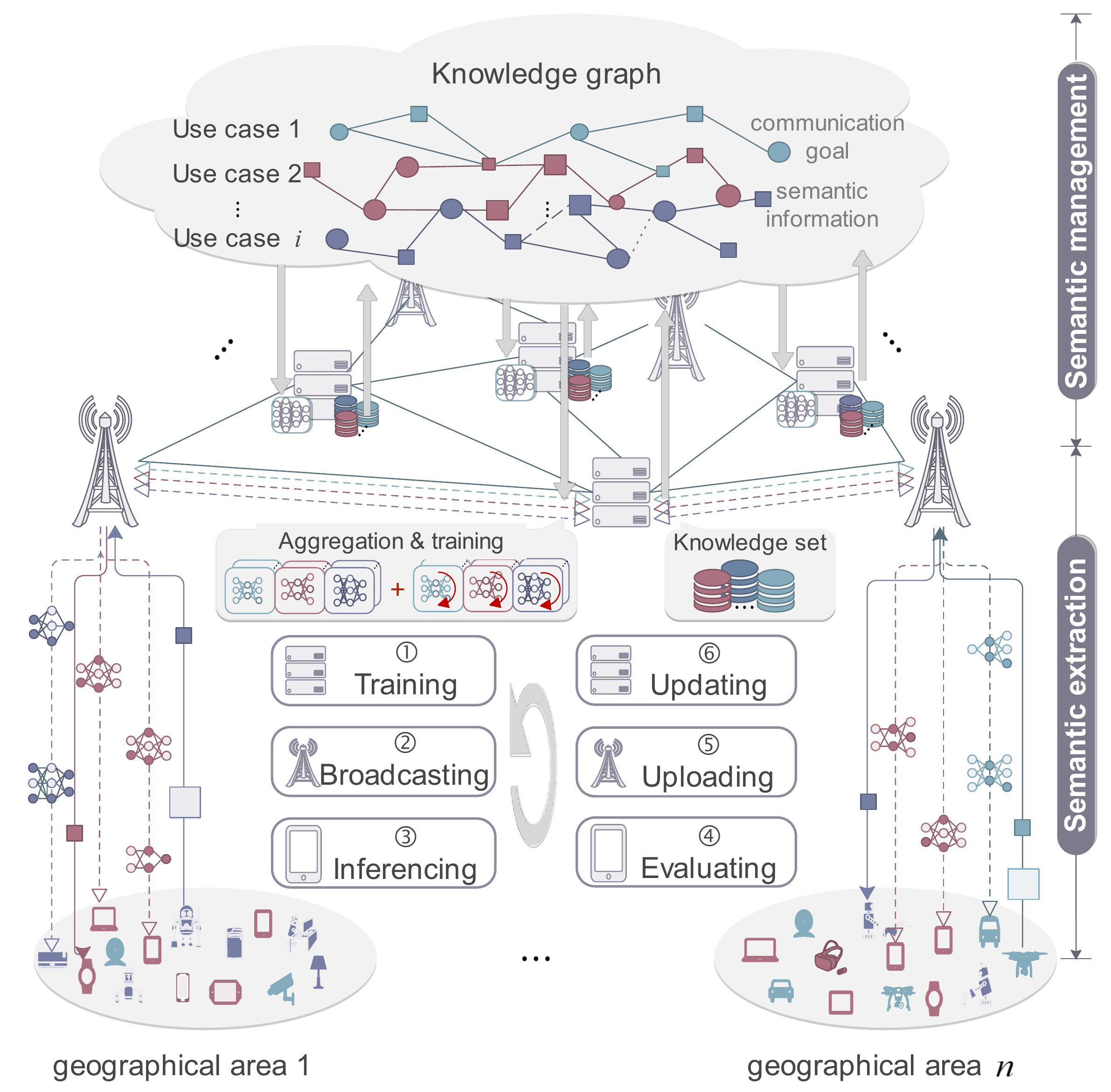}
  }
  \quad \quad
 \subfigure[]{
  \centering
  \includegraphics[scale = 0.35]{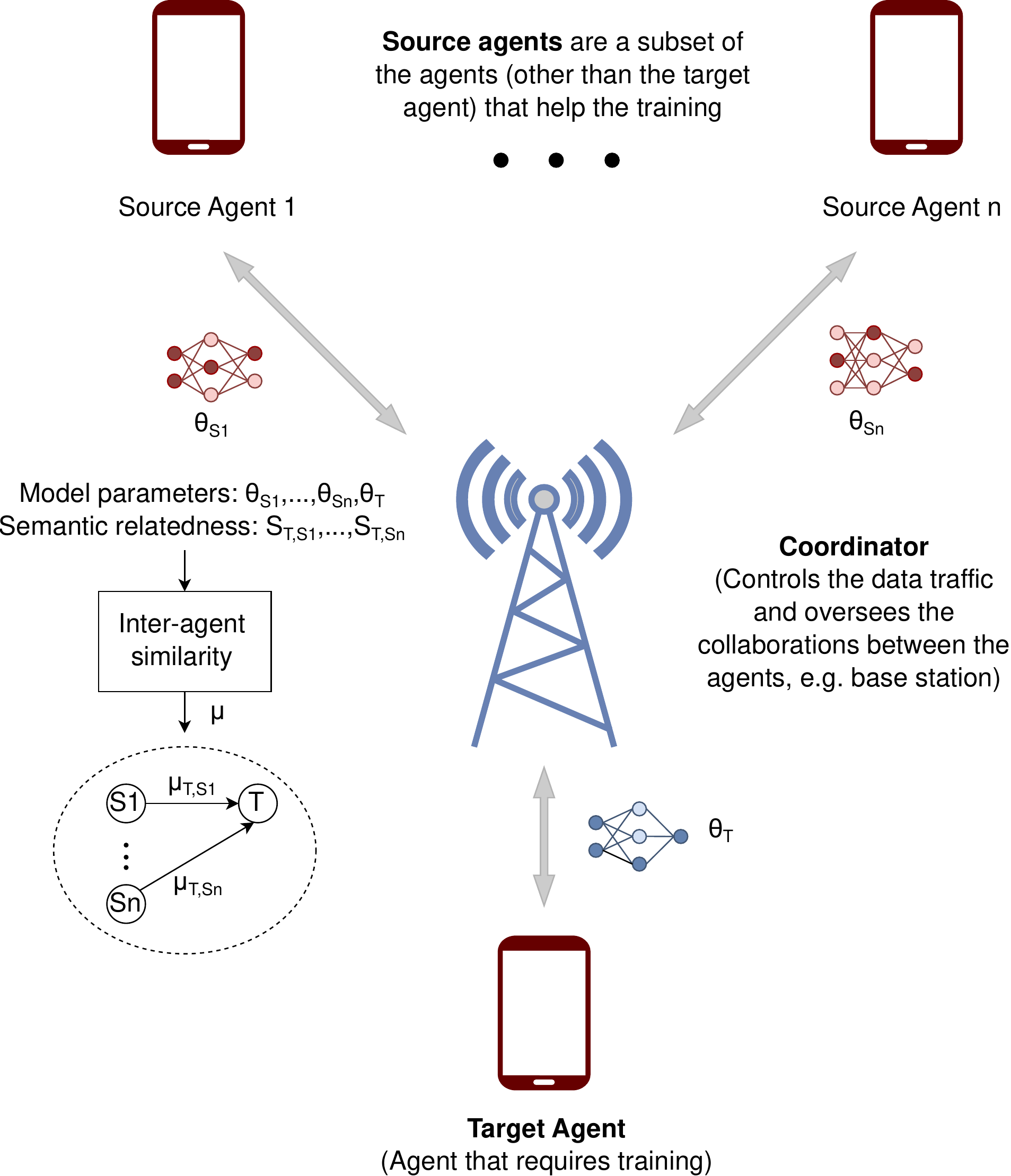}
  }

  \caption{ System models for the edge-enabled SemCom and Semantic-aware edge intelligence: a) architecture for edge-enabled SemCom; b) example structure of collaborative deep reinforcement learning.
  }
  \label{system}  
\end{figure*}
\subsection{Federated learning enabled SE}

In this subsection, we address the first two  challenges by integrating edge intelligence with SemCom. Given the powerful computation and caching capabilities of the edge servers, the background knowledge storage and  SE model training can be performed at the  edge. In this way, computation and communication latency for training (mentioned in the first challenge) can be reduced~\cite{xie2020lite}. Meanwhile, the edge server can serve as  an authoritative intermediary for knowledge sharing, thereby eliminating the need for all communication parties to  fully share each other's background knowledge~\cite{yang2021semantic}. This mitigates the second challenge.

We consider a common urban scenario shown in Fig.~\ref{system}(a). The end devices are typically clustered into different groups according to their associated access points or transmission requirements. Then, the edge server conducts SE model training and end devices employ well-trained model to perform SE. With Federated Learning (FL), the trained SE model parameters in edge servers can be exchanged directly with other edge servers with the identical tasks to accelerate the training process, thereby improving the generalization performance of the model in a privacy- preserving manner. The key procedures are outlined as below.
\begin{enumerate}[Step 1]
    \item Edge servers perform the pre-training or fine tuning for specific SE tasks based on each communication group’s shared background knowledge. Model parameter exchange and federated aggregation is performed over separate communication groups with the same communication goals but not a shared knowledge background.  (\textit{Edge server})
    \item  The derived global models are broadcast separately to each communication group. (\textit{Access point}) 
    \item The source devices generate the raw data. The destination devices receive SI. Then, the SE model is utilized to encode and decode SI. (\textit{End device})
    \item The destination devices evaluate the accuracy of SI during the communications for data labeling. (\textit{End device})
    \item The newly labeled SI  and/or corresponding raw data are uploaded to the edge servers. (\textit{Access point})
    \item The edge servers perform the regular updates for the knowledge sets according to uploaded  information and raw data for fine tuning of the SE model.  (\textit{Edge server})
\end{enumerate}

\subsection{Efficient semantic extraction based on edge-sharing knowledge-graph}
\label{sec:kg}

In this subsection, we focus on the third challenge. Inspired by the KB-assisted SE~\cite{yang2021semantic}, we propose to construct an edge-sharing knowledge graph (KG), which stores the underlaying
relations between communication goal and SI, for semantic management to enable more computationally-efficient SE.

 In general, a sophisticated KG heavily relies on a large deep learning model and a complete knowledge set. Fortunately, this is feasible in the framework of edge intelligence, where the KG can be cached at the edge servers and the available  related knowledge sets can be accessed at reduced link distances. Although the KG construction is also a computation-intensive task, the structure of KG is much more fixed than that of having to retrain separate SE models for various tasks. Once the KG is constructed, it can be cached at the edge servers to facilitate the computation-efficient SE. 

As an illustration, we consider the use case of KG towards SemCom in intelligent transportation networks. Since a well-trained convolutional neural network (CNN) for multiple object identification embeds all the feature maps related to different objects, the gradients of the output of the CNN with respect to feature maps can be treated as the importance weights of the feature map to different objects~\cite{yang2021semantic}. Accordingly, the KG can be established by storing the importance weights of all feature maps for the tasks with different identification targets~\cite{yang2021semantic}. In this sense, the SE for single object identification can be executed according to the important feature map, thereby avoiding the need for specialized training and also removing redundant details from the image for more efficient transmission. Meanwhile, although autonomous vehicles and unmanned
aerial vehicles (UAVs) work in distinct environments and have unique task specifications, they also share several similar characteristics and communication goals such as collision avoidance and path planning. Therefore, KG and transfer learning techniques (for the initialization of SE model parameters) can be applied to the training of SE to save much computation resource of the vehicles and UAVs.

\subsection{Research directions}
 While the above-mentioned network architecture can facilitate the development of efficient SemCom, there are still open issues to be solved before it can be implemented in practice, some of which are highlighted  below.
\begin{itemize}
    \item \textbf{Interpretability and explainability of SE}: As the unexpected information often appears in communications, the black-box nature of SE method hinders its implementation. Hence,  interpretability in SE needs to be studied to associate possible causes and results and to guide improvements to the SE model. Meanwhile, explainability in SE can identify the SI hidden in deep nets, which paves the way to the KG-based efficient SE across multiple modalities and tasks described in Section~\ref{sec:kg}. However, most existing SE methods are not explainable.
    
    \item\textbf{Semantic-noise based privacy preserving}:  For the communication groups with similar background knowledge and communication goals, eavesdropping becomes easy. Considering the success of covert communication in which artificial noise is introduced for  secure wireless transmissions, artificially increasing the mismatch to generate semantic noise may also serve as a potential method to enable secure SemCom. 
    
    \item \textbf{Variable length semantic encoding}: Existing works merely consider the dynamic channel gains in SE without the concern of resource constraints. However, in a multi-user scenario, the fluctuation in resources, such as available spectrum and transmit power, can have a non-negligible impact on the SemCom performance. The methods of achieving variable-length semantic encoding to cope with dynamic network resources remains thus to be an open research question.
\end{itemize}

\section{Semantic-aware Edge Intelligence}
There are many well-studied DL models that can be deployed at end devices to enable the intelligentization of edge networks. Moreover, intelligent agents have increasingly been deployed for edge orchestration. However, DL model optimization comes at the cost of bandwidth and energy resources. The situation is exacerbated by the rapid growth of edge intelligence networks. Besides the edge-enabled SemCom, we discuss in this section our approach to improve the performance and accuracy of DL enabled-edge intelligence networks under limited communication resources using semantic-aware methods. Specifically, we explore the impact of semantic awareness on several resource consuming tasks of edge intelligence networks, e.g., training and knowledge sharing among intelligent agents.

\subsection{Semantic-aware intelligent agent}
The ever increasing complexity of tasks in smart systems (e.g., autonomous driving, medical diagnosis, navigation)  calls for the adoption of intelligent agents that can learn and adapt based on their own experience. One of the promising deep learning based methods is DRL. For example, DRL-based methods are used in unmanned aerial vehicles (UAVs) to navigate large-scale complex environments, in robotics for labor-intensive tasks, and in SemCom for effective semantic extraction. With DRL, each intelligent agent will learn an optimal policy by maximizing a pre-defined task-oriented reward/return function. The learned policy is then used for decision making after the training, e.g., extracting useful semantic features in SemCom. In most smart services, intelligent agents are limited to the environment where they are deployed. As a result, the DRL model is trained by a limited dataset that does not fully represent the complex real-world environment. The monotonous experience induces overfitting issues, long convergence time, and sub-optimal performance of the DRL model.

To generalize the model experience, collaborative DRL (CDRL) is proposed to allow the agents to learn an optimal policy collaboratively by exchanging their model parameters or policies (Fig.~\ref{system}(b)). Following the idea of inter-agent knowledge sharing, \cite{yun2021attention} proposes a graph attention exchange network enabled by SemCom to share the local information between the DRL agents of UAVs. A brief structure of CDRL is shown in Fig. \ref{system}(b). In real-world applications, not all source agents can be selected for training due to the limited bandwidth. Hence, the allocation of the bandwidth and the agents' performance is often jointly optimized to maximize the utilities of the bandwidth. In short, the purpose of the joint optimization is to minimize the training loss of the target agent by selecting the source agents that are most helpful for the training of the target agent, all while meeting the resource limitation. However, it is challenging to filter the helpful source agents because the agents have different environments, tasks, and action spaces. To select the source agents, some works consider the structural similarity among these agents. A common method to obtain the structural similarity between the agents is by measuring the cosine similarity between the agents' model parameters. The agents with higher structural similarities are then selected for the collaborative learning. However, the structural similarity cannot capture the similarity of the underlying tasks of the agents, i.e., agents with similar model structures may not share a similar task and this may result in a poor collaboration. 

To this end, \cite{lotfi2021semantic} proposes to consider both structural similarity and semantic relatedness when selecting the source agents. To obtain the semantic relatedness, the target agent is trained for a fixed number of steps under the policy of the source agent, and the average return value is taken as the semantic relatedness. From the experiment results, it is shown that using the same bandwidth, the average return of the DRL agents is improved when the semantic aspects are considered, up to 83\% higher than the baseline methods.


\subsection{Semantic-aware distributed deep learning at wireless edge networks}
A drawback of CDRL and distributed deep learning for the optimization of edge intelligence networks is the high communication overheads incurred for the sharing and exchange of model parameters and policies. For example, while FL has been proposed to optimize aspects of 6G networks, e.g., to enable efficient edge caching and placement, the process of model parameter exchange is susceptible to communication stragglers. Therefore, finding an efficient way to compress the model parameters is essential to reduce the communication overhead. Two common solutions are gradient sparsification and model parameter pruning. In short, both methods extract a subset of the original model parameters to reduce the parameter size for transmission. 

Some studies proposed to consider the semantics or importance of the parameters during the data compression. For example, instead of random sparsification, \cite{sattler2019sparse} proposes to drop the gradients with lower magnitude and transmit the gradients with higher magnitude. The magnitude of the gradients signify the importance of the gradients, with higher gradients deemed to be more important for the weight updates. In \cite{sattler2019sparse}, the gradient estimates are sparsified at the transmitter and only the positions of the non-zero elements are sent to the receiver. 
To reduce the number of parameters during FL, pruning methods are commonly used for dropping a fraction of the model parameters to reduce the communication overheads. To identify the important parameters, \cite{jiang2019model} adopts adaptive model pruning where the importance of the model parameters are measured by their contribution to the future training.

\subsection{Research Directions}

Below, we highlight the open challenges for the SemCom enabled edge intelligence.

\begin{itemize}
    
    \item \textbf{DL-based SE for Task Similarity:} Although the proposed semantic relatedness metric in \cite{lotfi2021semantic} helps to improve the transmission performance, the extra training steps to obtain the return value will greatly reduce the system efficiency. Furthermore, it remains unclear how the number of training steps is determined, thereby limiting the scalability of this hand-crafted method. For future works, the semantic relatedness between the agents can be extracted using a deep learning network. The deep learning network can take the model parameters of the agent as input and output embeddings as the semantic representations of the agent. The network can be trained by minimizing the similarity between the output embeddings of different tasks, and maximizing the similarity of the output embeddings of the similar tasks. In this way, the semantic representation can be extracted to calculate the task similarity between the agents. As such, the most efficient source agents can be selected for the fixed bandwidth usage. Moreover, this method removes the need to arbitrarily define the number of training steps required.
    
    \item \textbf{Semantic Compression of Model Parameters:} The proposed methods for reducing the size of gradients and model parameters often require the setting of hyperparameters, e.g., degree of gradients sparsity. However, identifying the optimal values for the hyperparameters could be computationally inefficient. Recent works in deep learning enabled SemCom show that data transmission is more efficient and noise-tolerant by using the semantic encoding instead of conventional source coding. Future works can adopt semantic-aware model parameter exchange between the end devices. Such a system can follow the semantic encoder/decoder structure in \cite{xie2020lite} where the input parameters are first encoded by the transmitter using a semantic encoder and a channel encoder, before sending the encoded information to the receiver. The received signal is decoded by channel decoder and semantic decoder at the receiver to reconstruct the original data. For semantic text transmission, the SI sent by the transmitter carries the information used to reconstruct the text data at the receiver's end. In the case of CDRL, the encoded SI is the essential information to reconstruct the model parameter at the receiver's end.

\end{itemize}

\section{Case Study: Resource allocation for the convergence of SemCom and edge intelligence}

To facilitate the convergence of SemCom and edge intelligence, it is necessary to redesign the resource allocation policies. The reason is that while classical communication systems aim to improve communication efficiency in terms of reducing the BER or SER, SemCom aims to transmit the data relevant to the transmission goal. In other words, most existing resource allocation frameworks are designed to maximize throughput without considering the semantic importance of the bit flow, especially from different users. Moreover, an edge enabled SemCom system will have to involve entities with conflicting goals. For example, the transmitting end user aims to maximize the efficiency of the SE model while minimizing computation cost, whereas the edge server that charges the end user for its services aims to maximize revenue while reducing operational costs.

As a case study, we propose a system model in which energy-constrained IoT devices harvest the energy wirelessly for the purpose of text transmission \cite{liew2021economics}. Different from existing studies that maximize the bit transmission rate, our proposed framework aims to maximize the \textit{semantic performance} of the system. We consider a wireless powered communication network where there are a hybrid access point (HAP) and multiple wireless powered IoT devices. The IoT devices are equipped with semantic encoder/decoder to encode/decode SI from text data. For example, SI of a sentence with 32 words is encoded as a 2-dimensional matrix with size 32$\times$16, where 16 is the number of output dimension of the semantic features. As suggested in Section \ref{edge-enabled-semcom}, this is achieved through utilizing the trained SE models cached on edge servers.

\begin{figure}[t]
\centering
\subfigure[]{

  \centering
  \includegraphics[scale = 0.5]{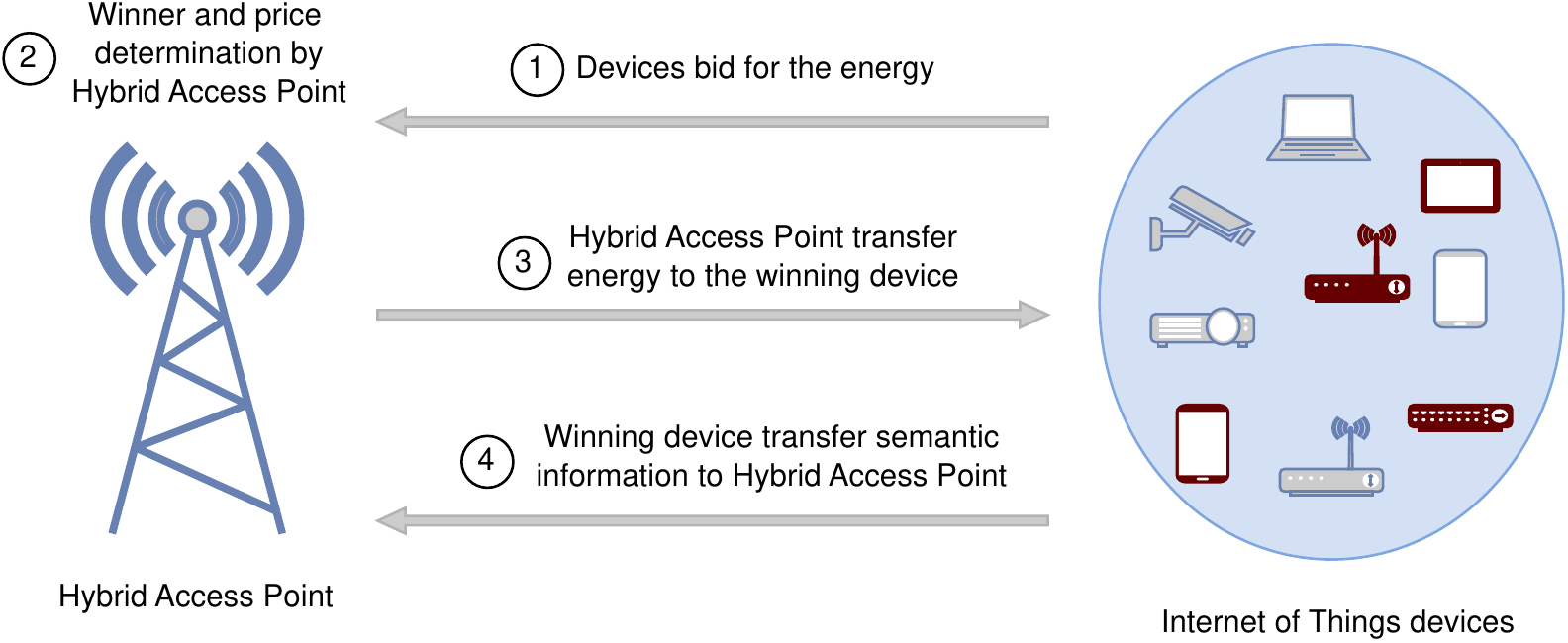}  
}

\subfigure[]{

  \centering
  \resizebox{200pt}{160pt}{%
  \begin{tikzpicture}
\begin{axis}[
    xlabel={Output Dimension $D$},
    xmin=1, xmax=16,
    ymin=0, ymax=0.9,
    legend style={fill=none, nodes={scale=0.8, transform shape}},
    xtick={1,2,3,4,5,6,7,8,9,10,11,12,13,14,15,16},
    ytick={0.1,0.2,0.3,0.4,0.5,0.6,0.7,0.8,0.9,1},
    legend pos=north west,
    ymajorgrids=true,
    xmajorgrids=true,
    grid style=dashed,
    legend cell align={left},
    every axis plot/.append style={very thick},
]

    \addplot[
    color=blue,
    ]
    coordinates {
(1,0.39550235)(2,0.40009948)(3,0.40945041)(4,0.41866887)(5,0.42247792)(6,0.42490115)(7,0.4295931)(8,0.43368545)(9,0.43733177)(10,0.4519554)(11,0.47728359)(12,0.51547686)(13,0.55437698)(14,0.61085957)(15,0.7460733)(16,0.86169747)
    };    \addlegendentry{similarity score};
    
    \addplot[
    color=red,
    dashed,
    ]
    coordinates {
(1,0.0944817)(2,0.09667912)(3,0.09386748)(4,0.10047062)(5,0.10116262)(6,0.10300542)(7,0.11076793)(8,0.11739845)(9,0.12781957)(10,0.15357989)(11,0.1940025)(12,0.27020956)(13,0.34242301)(14,0.44607532)(15,0.65054165)(16,0.82109432)

    }; \addlegendentry{1-gram BLEU score};
\end{axis}
\end{tikzpicture} } }

\subfigure[]{

  \centering
  \resizebox{200pt}{160pt}{%
  \begin{tikzpicture}
\begin{axis}[
    xlabel={No. of Iteration},
    ylabel={Revenue of HAP},
    xmin=1, xmax=2000,
    ymin=0.75, ymax=0.85,
    xtick={1,500,1000,1500,2000},
    ytick={0.75,0.8,0.85},
    legend pos=north east,
    legend style={fill=none, nodes={scale=0.8, transform shape}},
    ymajorgrids=true,
    xmajorgrids=true,
    grid style=dashed,
    legend cell align={left},
    every axis plot/.append style={very thick},
    /pgf/number format/.cd,
    1000 sep={},
]
\pgfplotstableread{train_res.txt}\resauction;
\addplot[
    color=red,
    dash pattern=on 6pt off 4pt on 2pt off 4pt,
    ]
    table
    [
    x expr=\thisrowno{0},
    y expr=\thisrowno{1}
    ] {\resauction};
    \addlegendentry{Deep Learning based Auction};

\addplot[
    color=blue,
    dashed,
    dash pattern= on 8pt off 4pt,
    ]
    table
    [
    x expr=\thisrowno{0},
    y expr=\thisrowno{3}
    ] {\resauction};
    \addlegendentry{Second-Price Auction};

\end{axis}
\end{tikzpicture}}
}
\caption{System model and experiment results of the case study: a) IoT devices bid for the energy from Hybrid Access Point; b) BLEU score and sentence similarity; c) revenue of Hybrid Access Point.}
\label{fig:casestudy}
\end{figure}
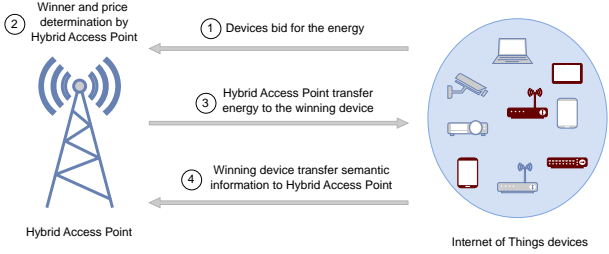
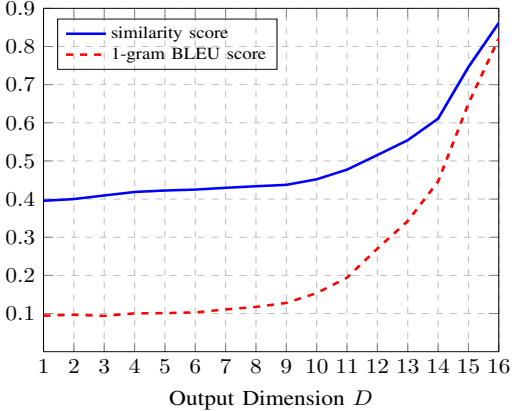
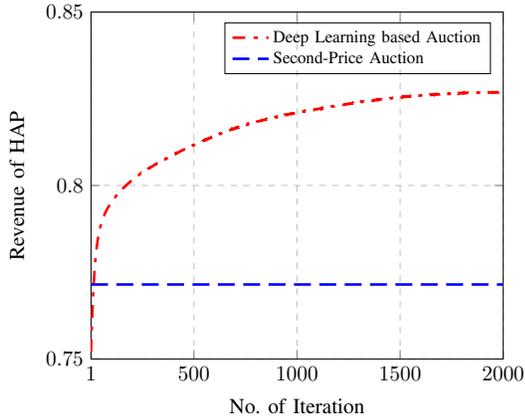

In the system (Fig. \ref{fig:casestudy}(a)), the HAP is considered to transmit energy to only one IoT device at a specific time. To decide the energy allocation, an auction mechanism is proposed where the IoT devices will bid for the energy, and the HAP will determine the winner and price. As the received energy is limited, some IoT devices need to reduce the output feature dimension to fit the energy budget. The sentence similarity and BLEU score under different output dimension is shown in Fig. \ref{fig:casestudy}(b). The number of bits that the devices can send upon receiving the energy is first obtained. Then, the IoT devices will adjust the output dimension to fit the data budget. As the objective of the transmission is to transfer SI, the IoT devices have more incentive to bid higher if they can achieve better semantic performance. The bids are derived from the sentence similarity and BLEU score (discussed in Table \ref{metrics}).  In general, the higher the sentence similarity and BLEU score, the higher the bid submitted by the devices. 

The winner and price are determined by a DL-based auction mechanism to maximize the revenue of the HAP. As shown in Fig. \ref{fig:casestudy}(c), the DL-based auction mechanism achieves higher revenue as compared to the traditional Second-Price Auction in which the highest bidder wins the energy allocation and pays the price of the second highest bidder. By maximizing the revenue of the access point, the price paid by the winning IoT device is also maximized. Hence, the energy is delivered to the device that values it the most (pay the maximized price) to ensure effective SemCom, all while fulfilling the desired properties of individual rationality and incentive compatibility for the auction. In the future, we can explore the semantic aware resource allocations for more data types, e.g., image transmission, video transmission, speech signal transmission.

\section{Conclusion }

In this article, we first provided a tutorial on SemCom. We discussed the SE techniques and performance indicators that vary from the CIT. Then, we motivated the edge-driven SemCom and the SemCom-driven edge, highlighting how the component of the two technologies can play an instrumental role towards the efficient intelligentization of future networks. We also discussed open research issues, as well as provide a case study of semantic-aware resource allocation.

\bibliographystyle{IEEEtran}
\bibliography{fl-uav}

\section*{Biographies}
\small
{WANTING YANG} received the
B.S. degree in 2018 from the Department of Communications
Engineering, Jilin University, Changchun,
China, where she is currently working toward the
Ph.D. degree with the College of Communication
Engineering. Her research interests include wireless
video transmission, learning, ultra-reliable, and low-latency
communications.

{ZI QIN LIEW} received the degree (Hons.) in electronic and electrical engineering from the Nanyang Technological University in 2018. He is currently an Alibaba Ph.D. Candidate with Alibaba Group and the Alibaba-NTU Joint Research Institute, Nanyang Technological University, Singapore. His research interests include wireless communications and resource allocation.

{WEI YANG BRYAN LIM} is currently pursuing the Ph.D. degree (Alibaba Talent Programme) with the Alibaba-NTU Joint Research Institute (JRI), Nanyang Technological University (NTU), Singapore. His research interests include edge intelligence and resource allocation.

{ZEHUI XIONG}  is an Assistant Professor at Singapore University of Technology and Design. Prior to that, he was a researcher with Alibaba-NTU Joint Research Institute, Singapore. He received the Ph.D. degree in Computer Science and Engineering at Nanyang Technological University, Singapore. He was a visiting scholar with Princeton University and University of Waterloo. His research interests include wireless communications, network games and economics, blockchain, and edge intelligence.

{DUSIT NIYATO} [IEEE Fellow] is currently a Professor with
the School of Computer Science and Engineering and, by courtesy,
School of Physical and Mathematical Sciences, Nanyang Technolog-
ical University, Singapore. He has published more than 380 technical
papers in the area of wireless and mobile networking, and is an
inventor of four U.S. and German patents. He was named the
2017–2021 Highly Cited Researcher in Computer Science. He is
currently the Editor-in-Chief for IEEE Communications Surveys and
Tutorials

{XUEFEN CHI}  is currently a Professor with the Department of Communications Engineering, Jilin University, Changchun, China.  She was a Visiting Scholar with  the School of Electronics and Computer Science, University of Southampton, Southampton, U.K., in 2015.  Her current research interests include machine type communications, indoor visible light communications, random access algorithms, delay-QoS guarantees, and queuing theory and its applications.

{XIANBIN CAO}   received the Ph.D. degree in signal and information processing from the University of Science and Technology of China, Hefei, China, in 1996. He is the Dean and a Professor with the School of Electronic and Information Engineering, Beihang University, Beijing, China. His research interests include intelligent transportation systems, airspace transportation management, and intelligent computation

{KHALED B. LETAIEF} [IEEE Fellow] received his Ph.D. degree from Purdue University. He has
been with HKUST since 1993 where he was the Dean of Engineering, and is now a Chair Professor and the New Bright Professor of Engineering. From 2015 to 2018, he was with HBKU
in Qatar as Provost. He is an ISI Highly Cited Researcher and
a recipient of many distinguished awards. He has served in
many IEEE leadership positions including ComSoc President,
Vice-President for Technical Activities, and Vice-President for
Conferences.

\end{document}